\documentclass[final,5p,times,authoryear]{elsarticle}

\usepackage{graphicx}
\usepackage{amsmath}
\usepackage{amssymb}
\usepackage{amsmath} 
\usepackage{soul,color}
\usepackage{placeins}
\usepackage{tikz}
\usepackage{booktabs}
\usepackage[caption=false, font=footnotesize]{subfig}
\usepackage{algorithm}
\usepackage{algpseudocode}

\usetikzlibrary{arrows.meta,positioning,calc}
\usepackage[most]{tcolorbox}

\newcommand \auxdRelgz	{ \kappa_4 }
\newcommand \auxRelcz		{ \kappa_1 }
\newcommand \auxRelgz		{ \kappa_3  }
\newcommand{\dtau}{\mathrm{d}\tau}
\newcommand \ksp  { k_\mathrm{s} }
\newcommand \lambdasat	{ \kappa_2 }
\newcommand \param		{ p } 
\newcommand \npar		{ q } 
\newcommand \parnor {\theta} 
\newcommand \parnornom {\parnor^\mathrm{nom}} 
\newcommand \Rel			{ \mathcal R }
\newcommand \Relcz		{ \kappa_1 }
\newcommand \Relgz		{ \kappa_3 }
\newcommand \tf			{ t_\mathrm{f} }
\newcommand \vcoil		{ u }
\newcommand \uff {\vcoil_{f{\!}f}}
\newcommand \vel			{ v }
\newcommand \vs         {\vspace{0pt}}
\newcommand \zsp  { z_\mathrm{s} }

\newcommand \g  { g }
\newcommand \X  { \varp }
\newcommand \tildeff  { (\tilde{J} - J^* ) }
\newcommand \varp  { X }
\newcommand \varpbest    {\varp^\mathrm{b}}
\newcommand \varpnom    {\varp^\mathrm{nom}}
\newcommand \e    {e}
\newcommand \s    {\deltaX^\mathrm{DF}}
\newcommand \deltaX    {\delta}

\let\originalleft\left
\let\originalright\right
\renewcommand{\left}{\mathopen{}\mathclose\bgroup\originalleft}
\renewcommand{\right}{\aftergroup\egroup\originalright}

\newcommand{\CommentLeft}[1]{\Statex \hspace{-0.2cm} \(\triangleright\) #1}

\journal{European Journal of Control}

\begin{document}

\begin{frontmatter}

\title{
A Hybrid Algorithm for Iterative Adaptation of Feedforward \\ Controllers: an Application on Electromechanical Switches
} 

\author{Eloy Serrano-Seco, Eduardo Moya-Lasheras, Edgar Ramirez-Laboreo}  

\affiliation{organization={Departamento de Informatica e Ingenieria de Sistemas (DIIS) and Instituto de Investigacion en Ingenieria de Aragon (I3A)}, 
            addressline={\\Universidad de Zaragoza}, 
            city={Zaragoza},
            postcode={50018}, 
            country={Spain}}

\begin{abstract}
Electromechanical switching devices such as relays, solenoid valves, and contactors offer several technical and economic advantages that make them widely used in industry. However, uncontrolled operations result in undesirable impact-related phenomena at the end of the stroke. As a solution, different soft-landing controls have been proposed. Among them, feedforward control with iterative techniques that adapt its parameters is a solution when real-time feedback is not available. However, these techniques typically require a large number of operations to converge or are computationally intensive, which limits a real implementation. In this paper, we present a new algorithm for the iterative adaptation that is able to eventually adapt the search coordinate system and to reduce the search dimensional size in order to accelerate convergence. Moreover, it automatically toggles between a derivative-free and a gradient-based method to balance exploration and exploitation. To demonstrate the high potential of the proposal, each novel part of the algorithm is compared with a state-of-the-art approach via simulation.
\end{abstract}


\end{frontmatter}

\section{Introduction}\label{sec:introduction}

Today, solenoid valves~\citep{angadi2022critical} and electromechanical relays~\citep{bojan2019design} are used in virtually all industries, ranging from household appliances and automotive applications to robotics and medical devices. In general, the basic operating principle of all electromechanical switching devices is similar: when electrical energy is applied, a magnetic force accelerates a moving component to the end of the stroke. This causes undesirable phenomena, including bouncing and violent impacts, which result in premature device wear and acoustic noise. In an effort to solve or reduce these phenomena, several control strategies have been proposed, generally with the same objective: to reach the final position with zero velocity. Among these strategies are those based on backstepping control~\citep{mohammad2022output}, sliding-mode control~\citep{deschaux2018nonlinear}, extremum-seeking adaptive control~\citep{Benosman2015}, or iterative learning control~\citep{moya2024iterative}.
 
Like other authors~\citep{braun2018observer}, some of our previous works employ a feedforward controller for two main reasons. First, a dynamic property of these systems, differential flatness, allows us to easily design the controller by model inversion. Secondly, feedforward control provides immediate responses to reference changes and is able to compensate for known disturbances. Despite its advantages, it alone is not robust to design errors, modeling errors, or system changes. To address these limitations, various complementary strategies exist, including conventional feedback controllers with observers \citep{schroedter2018flatness}, learning algorithms~\citep{grotjahn2002model}, and parameter adjustments based on measurable variables~\citep{yeh1999optimal}. Nevertheless, all the previously mentioned controllers are dependent on feedback of the variable to be controlled. However, in some cases, these variables cannot be measured or observed due to economic or technical constraints.

Therefore, we explored an alternative~\citep{moya2023IFAC} based on run-to-run controllers. The main idea is to iteratively update the feedforward controller parameters from measurable variables that, even though they cannot be directly used to observe and control the variable of interest, they provide a performance index for each iteration (i.e., switching operation). In a first approach, the iterative adaptation law was implemented using a Pattern Search~\citep{lewis2000pattern} algorithm. Although the method is computationally light and accurate, it requires too many evaluations to converge. Specifically, it needs 2$\,\npar\,+\,$1 evaluations (where $\npar$ is the search space dimension) to determine whether to move to a new point. Our latest work~\citep{edgar2024ECC} demonstrated that convergence speed can be improved through sensitivity-based parameter reduction. However, it did not offer an automated approach for determining the number of parameters to be reduced for a given problem, among other limitations. 

In this line, \citep{loshchilov2011adaptive} presents an Adaptive Coordinate Descent algorithm. The strategy involves periodically updating the coordinate system by a Covariance Matrix Adaptation Evolution Strategy (CMA-ES) and Adaptive Encoding to decompose the problem into as many one-dimensional problems as dimensions in the general problem. Although it is an interesting idea, as concluded by the authors in~\citep{hansen2001completely}, the function evaluations needed are about 10$\,\npar$, 30$\,\npar$ for a real-world search problem, and 100$\,\npar^2$ for complete adaptation. Given that each evaluation involves a switching operation, the number of switching operations with unsatisfactory performance would be excessively high.

In terms of one-dimensional search, the authors of~\citep{loshchilov2011adaptive} suggest derivative-free methods or the use of gradients. Gradient methods are a powerful tool, especially if the objective function is known. A similar alternative are subgradient methods, since they can work with approximations based on the value of the function to be optimized across the search space. Another option are the sign gradient descent methods, first introduced in the RProp (Resilient Propagation) algorithm~\citep{moulay2019properties}. Despite being technically a gradient-based method, the RProp algorithm has a low computational load, as it only needs to calculate the sign of the gradient, not the gradient itself. Nevertheless, adjusting the hyperparameters of these algorithms can be a challenging task. In contrast, some gradient descent methods implement an adaptive step size without the need for hyperparameters.

This paper presents a new Run-to-Run controller based on an Adaptive Coordinates algorithm (R2R-AC) in order to automate the improvement process described in~\citep{edgar2024ECC} and to enhance the performance of the iterative adaptation law. This new algorithm leverages the controller sensitivity with respect to its parameters to calculate an alternative search basis that decomposes the initial $\npar$-dimensional problem into $\npar$ one-dimensional problems to optimize on the descending coordinate with the highest improvement potential. We analyze and compare three versions of the algorithm that differ in how the step size is computed: one based on derivative-free methods, another based on \mbox{gradient-based} algorithms, and a hybrid one that toggles between the other two to enhance the exploration-exploitation tradeoff.

The paper is organized as follows. Section~\ref{sec:background} provides a concise overview of the dynamic and control model that has prompted the development of the proposed algorithm. Section~\ref{sec:algorithm} develops the iterative adaptation law of the R2R-AC strategy and discusses the different versions previously mentioned. Section~\ref{sec:simresult} contains simulation results that demonstrate the functionality of our proposals and the comparison with a state-of-the-art feedforward run-to-run controller. Finally, the conclusions are discussed in Section~\ref{sec:conclusions}.

\vs
\section{Background of the control system} \label{sec:background}

In this section, we briefly describe the system dynamics and control where the need for the proposed new algorithm has arisen. For a more detailed explanation, readers are referred to our works~\citep{moya2023IFAC} and~\citep{edgar2024ECC}.

\subsection{System dynamics} \label{sec:system}

The system is modeled as a single-coil reluctance actuator, subject to two types of forces: passive elastic forces—typically modeled as ideal springs—and a magnetic force. The magnetic force is generated when current flows through the coil, causing an inner fixed core to become magnetized and attract the movable core. The typical method of supplying the actuator with power is by providing a voltage. We describe the dynamics of the system using a state-space model, where the voltage $u$ is the input to our system, and the position $z$, velocity $\vel$ and magnetic flux linkage $\lambda$ are the state variables. The state equations are defined as
\begin{align}
    \dot z &= \vel, \label{eq:dynz} \\
    \dot \vel &= \frac{1}{m}\, \left( -k_\mathrm{s} \, \left(z-z_\mathrm{s}\right) - \frac{1}{2}\,\lambda^2 \, \frac{\partial \Rel}{\partial z} \right), \label{eq:dyndz} \\  
    \dot \lambda &= -R \, \lambda\, \Rel(z,\lambda) + u, \label{eq:dynlambda}
\end{align}
where $m$, $k_\mathrm{s}$, $z_s$, $R$, and $\mathcal R$ are the moving mass, the spring stiffness, the spring resting position, the coil resistance, and an auxiliary function based on the magnetic reluctance concept, respectively. This auxiliary function considers the magnetic saturation and flux fringing phenomena,
\begin{equation}\label{eq:Rel}
    \Rel(z,\lambda) =  \frac{\Relcz}{1-|\lambda|/\lambdasat} + \Relgz + \frac{\auxdRelgz \, z}{1 + \kappa_5\, z \, \log(\kappa_6/z)},
\end{equation}
where $\auxRelcz$, $\lambdasat$, $\auxRelgz$, $\auxdRelgz$, $\kappa_5$, and $\kappa_6$ are positive constants. Overall, the system dynamics depends on $q=9$ uncertain parameters, which can be grouped in the parameter vector $p$.
\begin{equation}\label{eq:param0}
        \param = \\ \left[
        \,\, k_\mathrm{s} \,\,\,\, z_\mathrm{s} \,\,\,\, m \,\,\,\, \auxRelcz \,\,\,\, \lambdasat \,\,\,\, \auxRelgz \,\,\,\, \auxdRelgz \,\,\,\, \kappa_5 \,\,\,\, \kappa_6 \,\, 
    \right]^\intercal .
\end{equation}
Note that the resistance $R$ is treated independently as a parameter without uncertainty, as it can be easily measured.

\subsection{Control} \label{sec:control}

\begin{figure}[t]
    \centering
	\def\sumoffset{1mm}
	\def\sumoffsetaux{1.5mm}
	\def\nodex{10mm}
	\def\nodey{13mm}
	\def\arrowsep{3mm}
	\def\lwt{0.3mm}
	\def\lwn{0.4mm}
	\def\lwd{0.5mm}
	\begin{tikzpicture}[
		node distance = \nodey and \nodex,
		box/.style = {draw, minimum height=10mm, minimum width=12mm, align=center},
		sum/.style = {circle_, draw, node contents={}},
		>={Stealth[width=2mm,length=3mm]}
		]
		\node (ref) [] {$z(t)$};
		\node (ff) [box, right=of ref, xshift=-5mm] {Feedforward\\controller};
 		\coordinate[right=of ff,xshift=90] (c2);
		\node (plant) [box, right=of ff,xshift=15mm] {System};
		\coordinate[below=of plant.center, yshift=-0.5*\arrowsep] (c5);
		\coordinate[above=of plant.center] (c7);
		\coordinate[above=of c7] (c8);
		\coordinate[above=of c8] (c9);
		\coordinate[above=of ff.center] (c10);
		\node (cost) [box, above=of plant.center,yshift=20, anchor=center] {Cost};
		\node (opt) [box, above=of ff.center,yshift=20, anchor=center] {Iterative\\adaptation law};
		\draw[->,line width=\lwt] (ff) -- node[above] {$\uff(t,\parnor_{k})$} (plant);
		\draw[dashed,->,line width=\lwn] (plant) -- node[left] {$y(t)$} (c7) -- (cost);
		\draw[dashed,->,line width=\lwn] (cost) -- node[above] {$J_{k}$} (opt);
		\draw[dashed,->,line width=\lwn] (opt) -- node[left] {$\parnor_{k+1}$} (ff);
		\draw[->,line width=\lwt] (ref) -- (ff);
	\end{tikzpicture}
    \caption{General control diagram. The subscript $k$ denotes the variables of the $k$-th evaluation of the run-to-run adaptation law. The feedforward block computes $\uff$ from the parameter vector $\parnor$ and the desired reference signal~$r$. The adaptation law updates the feedforward parameters $\parnor$ using the cost $J$, which is derived from the measurable output~$y$.}
	\label{fig:v_ctrl_diag}
 \vspace{-3mm}
\end{figure}

The control structure used in this study is schematized in Fig.~\ref{fig:v_ctrl_diag}. This is an iterative control design for a real-world scenario with two particularities:
\begin{itemize}
    \item 
    The variable to be controlled, the position $z$, cannot be fed back for several reasons.
    Firstly, a position sensor is more expensive than the switching devices. Secondly, a protective housing impedes access to the component whose position needs to be known. In addition, a real-time estimate is also unavailable.
    \item Errors in the model parameters are not negligible. These devices are produced at a low cost, with relaxed manufacturing tolerances, causing variability in the value of the parameters.
\end{itemize}
Due to the first point, the control strategy is focused on a feedforward controller. It is designed by model inversion, taking advantage of a structural property shared by such devices: differential flatness~\citep{levine2011}. Considering that the objective is soft-landing control, as in our previous works, $z(t)$ is the desired trajectory, designed as a $5$th-degree polynomial from $z_0$, the initial mechanical limit of motion of the moving component to be controlled, to $z_\mathrm{f}$, the final limit. The trajectory is defined over the time interval $[t_0, \tf]$, subject to boundary conditions of zero initial and final velocities and accelerations. In short, the feedforward control term, $\vcoil_{ff}$, is defined as a function of $z(t)$, its derivatives, and the parameter vector $\param$. Although this applies to a specific case, it can be generalized as $\vcoil_{ff} = \uff\left(t, \parnor \right)$ for parametric controllers where $\parnor$ represents any vector of control parameters. In our case $\parnor$ is a normalized version of $\param$.

Due to the second point, including a feedback loop to adapt the control parameters $\parnor$ is essential. Some previous works propose associating a measurement related to the control objective to a cost value $J$. In a real-world application, due to measurement difficulties, $J$ may be computed based on indirect measurements associated with the impacts \citep{moya2023IFAC,winkel2023reducing}. In simulation, however, the impact velocity $v_\mathrm{c}$ could be directly used as feedback,
\begin{equation}\label{eq:cost}
    J= \left\lvert {v_\mathrm{c}}\right\rvert.
\end{equation}
At this point, it is reasonable to assume that the process which relates $\parnor$ to $J$ is unknown or difficult to work with analytically. In response to this, the proposal is to use a black-box optimizer as the iterative adaptation law to minimize $J$ along switchings. Since we are focusing on a real-world application, the optimizer must be not only accurate, but also computationally light and able to quickly discard poorly performing evaluations to minimize unsatisfactory switching operations. Therefore, to speed up convergence, a state-of-the-art feedforward run-to-run controller based on a Pattern Search algorithm~\citep{edgar2024ECC} shows a strategy to reduce  the number of search dimension, without relying on further cost function evaluations. Assuming that larger changes in the control action translate into larger changes in the cost value, this strategy is based on a local sensitivity analysis of smooth (differentiable) controllers. The Fisher matrix, $\mathcal{F}({\parnor})$, which can be computed from the sensitivity of the controller with respect to $\parnor$,
\begin{equation} \label{eq:Fisher}
    \mathcal{F}({\parnor}) = {\int_{t_0}^{\tf} \Big[\Big(\dfrac{\partial \uff(t,\parnor)}{\partial \parnor} \Big)^{\intercal}\, \Big(\dfrac{\partial \uff(t,\parnor)}{\partial \parnor}\Big)\Big] \,\dtau}, 
\end{equation}
is evaluated at a nominal point $\parnornom$. Since $\mathcal{F}(\parnornom)$ is symmetric and positive semidefinite, the eigendecomposition and the singular value decomposition coincide. That is, the Fisher matrix can be expressed as \mbox{$\mathcal{F}({\parnornom}) = V \, \Lambda \, V^\intercal$}, where $V\in \mathbb{R}^{\npar\times\npar}$ is the orthonormal matrix with the eigenvectors (or singular vectors) of $\mathcal{F}({\parnornom})$ as columns and $\Lambda\in \mathbb{R}^{\npar\times\npar}$ is the diagonal matrix of the corresponding eigenvalues (or singular values). 

Finally, a transformation of the known vector $\parnor$ to a new vector $\varp\in \mathbb{R}^\npar$ is defined in terms of the basis change matrix $V$ as
\begin{equation} \label{eq:varphi_theta}
   \parnor = \parnornom + V (\varp - \varpnom) \ \Longleftrightarrow \ \varp = \varpnom + V^\intercal (\parnor - \parnornom),
\end{equation}
where $\varpnom$ is the nominal value of $\varp$, which can be chosen arbitrarily.

Thanks to this procedure, the controller is parameterized in such a way that the correlation between the sensitivities with respect to the new parameters in vector $\varp$ is low. In addition, $\Lambda$ provides information about these sensitivities, enabling the exclusion of coordinates with negligible sensitivity from the optimization process.

\vs
\section{New algorithm} \label{sec:algorithm}

This section is divided into two parts. The first one, following the idea presented in~\citep{loshchilov2011adaptive}, introduces a new algorithm which solves the open questions posed in~\citep{edgar2024ECC}, i.e., what is the appropriate number of dimensions to optimize in each situation and, since the analysis is local, how often the alternative coordinate system should be recalculated. The second part presents a classical \mbox{derivative-free} method, a \mbox{gradient-based} method, and a hybrid methodology that enables toggling between the two options to calculate the step size of the previous algorithm.

\subsection{Iterative adaptation law of the R2R-AC } \label{sec:mov}

The behavior of the algorithm is presented in Algorithm~\ref{alg:ADC}. The new algorithm must fulfill two requirements: it must eventually upgrade the alternative coordinate system and it must select automatically the number of search dimensions. We propose to convert the complete optimization of the \mbox{$\parnor$-problem} into successive \mbox{$\varp$-optimization} problems. Each \mbox{$\varp$-optimization} problem optimizes $\varp$ in a new canonical basis initialized at $\varpnom = \mathbf{0}_q$. This decision simplifies the transformation of $\varp$ into $\parnor$ as
\begin{equation} \label{eq:theta_X}
    \parnor = \parnornom + V \varp,
\end{equation}
with the only eventual needs to update $\parnornom$ as the best evaluated $\parnor$ (Algorithm~\ref{alg:ADC}, line~\ref{Alg1:line:reset}) and update the transformation matrix $V$ (line~\ref{Alg1:line:up_V}) as shown in the previous section (according to~\citep{edgar2024ECC}). From now on, for clarity, since we work mainly in the \mbox{$\varp$-optimization} space, with a slight abuse of notation, we denote the cost --- obtained for a value $\parnor$ that depends on $\varp$, \eqref{eq:theta_X} --- as a direct function of $\varp$: $J(\varp)=J$.

Once the strategy for updating the alternative coordinate system has been selected, the remaining tasks are to determine the timing of the update event and the search problem dimensions. The proposed solution is to first perform an exploration process to find the coordinate with the greatest descent and then to exploit this coordinate. For the exploration process, $V$ should be ordered such that the associated eigenvalues are arranged from largest to smallest, i.e., 
\begin{equation}
    \Lambda_{(1,1)} \geq \Lambda_{(2,2)} \geq ... \geq \Lambda_{(\npar,\npar)} ,
\end{equation}
where the subscripts in parentheses represent the position in the matrix. This step permits the ordering of the coordinates from the highest to the lowest sensitivity of the control action to the parameters $\varp$. The proposal for selecting the coordinate of greatest descent is based on pattern search methods: the origin of the coordinate system (line~\ref{Alg1:line:ev_best}) and two points along each coordinate (line~\ref{Alg1:line:ev_X}), $\varp^{+}$ and $\varp^{-}$, symmetrically located at a distance $\deltaX$ from the origin coordinate, are evaluated,
\begin{align}
    \varp^{+} &\leftarrow \varpbest + \deltaX \cdot \e_{d}, \label{eq:X+}\\
    \varp^{-} &\leftarrow \varpbest - \deltaX \cdot \e_{d}, \label{eq:X-}
\end{align}
where $\varpbest$ is the point associated with the lowest cost value, whose value is $\mathbf{0}_{\npar}$ at the start of each \mbox{$\varp$-optimization} problem, $\deltaX \in \mathbb{R}$ is the step size, and $\e_{d} \in \mathbb{R}^\npar$ is the unit vector that defines the direction of the $d \in [1, \npar]$ coordinate of the canonical basis. This pattern is evaluated sequentially coordinate by coordinate (lines~\ref{Alg1:line:begin_pattern}--\ref{Alg1:line:end_pattern}) until a cost-improving coordinate has been found. At this moment the exploration process ends and the exploitation process begins by a line-search (lines~\ref{Alg1:line:begin_line}--\ref{Alg1:line:end_line}). The algorithm continues to look for lower cost points along the corresponding direction and orientation by a method that embraces the philosophy of sign gradient descent algorithms,
\begin{equation}
    \varp^\mathrm{next} \leftarrow \varpbest + \text{sgn}\left(\varpbest_{(d)}\right) \, \deltaX \cdot \e_{d}, \label{eq:Xnext}
\end{equation}
where $\varp^\mathrm{next}$ is the next point to be evaluated, $\text{sgn}(\cdot)$ is the sign operator, and $\varpbest_{(d)}$ is the \mbox{$d$-th} component of $\varpbest$. Note that {\eqref{eq:Xnext}} only applies after a better point has been found, so in this case, $\varpbest_{(d)} \neq 0$.

When the evaluation of $\varp^\mathrm{next}$ does not improve the cost, the present \mbox{$\varp$-optimization} problem ends, $\parnornom$ and $V$ are updated and the \mbox{$\varp$-optimization} problem is reset (line~\ref{Alg1:line:reset}). Thus, it is not necessary to complete the pattern to shift it, and the number of dimensions of the problem is automatically reduced to the minimum allowed to obtain improvements. An exception applies to this sequence, when the algorithm has moved on the first coordinate (line~\ref{Alg1:line:if_d1}), $\parnor$ is not updated, so neither is $V$, and the evaluation of the second coordinate continues around the best point found on the first coordinate. The main reason is not to transform the algorithm into a single gradient search method in which the descending coordinate is calculated through the coordinate that further modifies $\uff$, because the convergence speed may decrease due to limited information and reduced opportunities to directly identify a new best point.

\begin{algorithm}[t]
\caption{Iterative adaptation law of the R2R-AC}\label{alg:ADC}
\begin{algorithmic}[1]
\Require{$\delta$, \hspace{1mm} $\parnornom$}, \hspace{1mm} $d \leftarrow 0$, \hspace{1mm}  $\varpbest \leftarrow \mathbf{0}_{\npar}$, \hspace{1mm} $J(\varp^\mathrm{next}) \leftarrow \infty$
\While{true}
\CommentLeft{Alternative coordinate system}
    \State $V \leftarrow$ sorted eigenvectors of $\mathcal{F}(\parnornom)$   \label{Alg1:line:up_V}  
\CommentLeft{Best descent coordinate exploration}
    \State Evaluate $J(\varpbest)$   
    \While{$J(\varp^\mathrm{next}) > J(\varpbest)$}  \label{Alg1:line:begin_pattern}  
        \State $d \leftarrow (d \bmod \npar)\,+\,1$     \Comment{ Next $d$ }
        \State $\varp^{+} \leftarrow \varpbest + \deltaX \cdot \e_{d}$\,; \hspace{5mm} $\varp^{-} \leftarrow \varpbest - \deltaX \cdot \e_{d}$
        \State Evaluate $J(\varp^{+})$ and $J(\varp^{-})$  \label{Alg1:line:ev_X} 
        \State $\varp^\mathrm{next} \leftarrow \arg \min_{\X \in \{\varp^{+},\,\varp^{-}\}}  J(\varp)$
        \State Update $\delta$    \label{Alg1:line:up_delta1}  \Comment{Algorithm~\ref{alg:dx}}
    \EndWhile  \label{Alg1:line:end_pattern}  
\CommentLeft{Descent coordinate exploitation: Line-search}
    \While{$J(\varp^\mathrm{next}) < J(\varpbest)$}   \label{Alg1:line:begin_line}  
        \State $\varpbest \leftarrow \varp^\mathrm{next}$
        \State $\varp^\mathrm{next} \leftarrow \varpbest + \text{sgn}\left(\varpbest_{(d)}\right) \, \deltaX \cdot \e_{d},$   \label{Alg1:line:ev_best}
        \State Evaluate $J(\varp^\mathrm{next})$  
        \State Update $\delta$     \label{Alg1:line:up_delta2}       \Comment{Algorithm~\ref{alg:dx}}
    \EndWhile   \label{Alg1:line:end_line}  
\CommentLeft{Reset the $\varp$ optimization problem}
    \If{$d \neq$ 1}     \label{Alg1:line:if_d1}  
        \State $\parnornom \leftarrow \parnornom + V \varpbest$\,; \hspace{2.5mm} $d \leftarrow 0$\,;  \hspace{2.5mm}  $\varpbest \leftarrow \mathbf{0}_{\npar}$   \label{Alg1:line:reset}  
    \EndIf
\EndWhile
\end{algorithmic}
\end{algorithm} 

To complete the algorithm, it only remains to define how to upgrade the step size (lines~\ref{Alg1:line:up_delta1} and \ref{Alg1:line:up_delta2}). This is discussed in the following subsection.

\subsection{Step-size update: hybrid method} \label{sec:step-size}

The new algorithm enables the $\npar$-dimensional problem to be reduced to the behavior of a one-dimensional problem. For the sake of simplicity, this subsection assumes that a descending shift always occurs. For clarity, we work with $\hat{\e}_{d}$, the unit vector of the desired direction and orientation.

The fact that there is no analytical information on the relationship between the cost value $J$ and the parameters to optimize $\varp$, together with the need of a computationally light and fast convergence process, has led us to the use of direct search methods based on updating a step size $\deltaX$. We explore two approaches: derivative-free methods and adaptive methods based on the objective function.

In the basic derivative method, the step size decreases as the number of evaluations increases; however, other methods allow for expanding the step size if certain conditions are met. The latter methods are used in algorithms as Pattern Search and RProp, among others. When it appears the process is progressing in the right direction, i.e., the cost improves, the step size should be increased (expansion) to reach the possibly distant optimum point more quickly. Conversely, when the process has reached a minimum, i.e., the cost does not improve, the step size should be decreased (contraction) to allow for an approach to the minimum cost and reduce fluctuations. In short, the next point to evaluate $\X^\mathrm{next}$ can be calculated as
\begin{equation}
     \X^\mathrm{next} = \X^\mathrm{b} + \s\cdot\hat{\e}_{d}, \label{eq:PS} 
\end{equation}
where $\X^\mathrm{b}$ is the point considered best, $\s$ is the \mbox{derivative-free} step size, which is eventually updated as
\begin{equation}
    \s \leftarrow   
    \renewcommand{\arraystretch}{1.4} 
            \Biggl\{\begin{array}{cc}
            \max(\alpha_\mathrm{con} \, \s, \, \s_\mathrm{min}) & \mathrm{if} \, J(\X^\mathrm{next}) > J(\X^\mathrm{b}) \\
            \min(\alpha_\mathrm{exp} \, \s, \, \s_\mathrm{max}) & \mathrm{if} \, J(\X^\mathrm{next}) \leq J(\X^\mathrm{b})
                        \end{array}, \label{eq:step_DF} 
\end{equation}
where $\alpha_\mathrm{con}$ and $\alpha_\mathrm{exp}$ are the contraction and expansion constants such that \mbox{$0\,<\,\alpha_\mathrm{con}\,<\,1\,<\,\alpha_\mathrm{exp}$}, and $\s_\mathrm{min}$ and $\s_\mathrm{max}$ are the minimum and maximum allowed step sizes, respectively.  Unfortunately, the tuning of $\alpha$ values suffers from a trade-off between faster convergence and the likelihood of reaching a local minima, which tends to be higher as the dimensionality of the problem increases.

\begin{figure}[t]
    \centering
    \includegraphics[trim={0 1mm 0 0},clip]{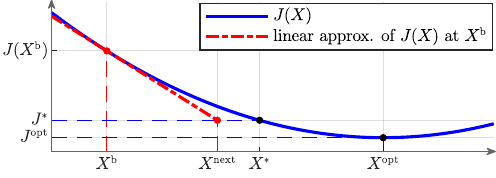}
     \vspace{-2mm}
    \caption{Geometric interpretation of the first-order method adopted}
    \label{fig:grad_exp}
    \vspace{-5mm}
\end{figure}

In contrast, adaptive methods based on the objective function are able to dynamically fit the step size, often utilizing gradient information in first-order methods. A typical geometric interpretation of these methods is illustrated in Fig.~\ref{fig:grad_exp} for an ideal situation with a convex objective function. The aim is to reach a point $\X^*$ with a lower cost $J^*$ from a known point $\X^\mathrm{b}$, for our algorithm, the best evaluated point. To this end, the step size is approximated as the ratio of the cost difference $J(\X^\mathrm{b})\,-\,J^*$ to the gradient $\g$ at $\X^\mathrm{b}$. To determine $J^*$, common approaches include treating it as a constant equal to $J^\mathrm{opt}$, the cost of the optimal point $\X^\mathrm{opt}$, i.e., the minimum value of the objective function. Even if the real value is unknown, $J^* = 0$ is adequate on many situations, but it is usually a strong assumption. Alternatively, $J^*$ can vary by iteration (Algorithm~\ref{alg:dx}, line~\ref{Alg2:line:up_J*}), which is often more practical, as large step sizes can be more detrimental than an advantage, leading to divergence. As shown in Fig.~\ref{fig:grad_exp}, larger step sizes introduce greater approximation errors, potentially affecting convergence.

\begin{algorithm}[]
\caption{Process to update the step size $\deltaX$}\label{alg:dx}
\begin{algorithmic}[1]
\Require{$\tilde{J}$,  $\tilde{\g}$,  $\beta$, $\s$, $\alpha_\mathrm{con}$, $\alpha_\mathrm{exp}$,  $\s_\mathrm{min}$,  $\s_\mathrm{max}$, $J^*$}
\For{$k \leftarrow 1$ to num. evaluations}
    \State Update $\tilde{J}$ and $\tilde{g}$       \Comment{\eqref{eq:up_J} and \eqref{eq:up_g}}
    \If{$\deltaX$ must be updated}
        \State Update $\s$ and $\deltaX^\mathrm{GB}$      \Comment{\eqref{eq:step_DF} and \eqref{eq:step_GB}}
        \State Update $J^*$     \label{Alg2:line:up_J*}
        \State $\deltaX \leftarrow   
                        \Biggl\{\begin{array}{cc}
                        \s & \mathrm{if} \, \tilde{J} \leq J^{*} \\
                        \deltaX^\mathrm{GB} & \mathrm{if} \, \tilde{J} > J^{*}
                        \end{array}$        \label{Alg2:line:up_J}
    \EndIf
\EndFor
\end{algorithmic}
\end{algorithm}

For our problem, our non-static search coordinate system complicates the collection of information on $\g$, as a reminder, the gradient with respect to $X$.  Thus, we propose a reinterpretation of this method by working with $\tilde{\g}$, an average slope of the trajectory followed by the algorithm that represents the average improvement capacity. In addiction, given the possible non-smoothed decreasing cost, we work with $\tilde{J}$, an average value of the cost. These are computed as
\begin{gather}
    \tilde{J} \leftarrow \beta \, \tilde{J} + (1 - \beta) J(\X_{k}), \label{eq:up_J}\\
    \tilde{\g} \leftarrow \sqrt{\beta \, \tilde{\g}^2 + (1 - \beta) \left(\frac{J(\X_{k})-J(\X_{k-1})}{||\varp_{k}-\varp_{k-1}||}\right)^2}, \label{eq:up_g}
\end{gather}
where $\beta < 1$ is a positive constant that acts as a decay factor, and the subscript $k$ refers to the evaluation number. With these adjustments, the next point to be evaluated and the value of the \mbox{gradient-based} step size $\delta^\mathrm{GB}$ are calculated, respectively, as
\begin{gather}
    \X^\mathrm{next} = \X^\mathrm{b} + \deltaX^\mathrm{GB}\cdot\hat{\e}_{d},      \label{eq:subgradient}\\ 
    \deltaX^\mathrm{GB} = \frac{\tildeff}{\tilde{\g}}.     \label{eq:step_GB}
\end{gather}
Analyzing the equation, an additional advantage of slope filtering is that it mitigates the problem of excessively oscillating step sizes. This concept is also used in several stochastic gradient descent algorithms, including RMSProp~\citep{ma2022qualitative}. Despite the adaptability of \mbox{gradient-based} methods, they are highly susceptible to the geometry of the objective function and the initial evaluations.

Our hybrid version for calculating $\deltaX$ tries to take advantage of the good performances of both strategies and avoid their drawbacks by toggling between~\eqref{eq:step_DF} and~\eqref{eq:step_GB} based on whether $\tilde{J}$ is less than or greater than $J^{*}$, respectively (line~\ref{Alg2:line:up_J}). In this form the first resource of the algorithm is~\eqref{eq:step_DF}, but, when convergence slows significantly or the evaluated point is far from the optimal point, the strategy toggles to \eqref{eq:step_GB}. The update of the step size $\deltaX$ is summarized in Algorithm~\ref{alg:dx}.

\vs
\section{Simulated results} \label{sec:simresult}

\begin{table}[t]
    \begin{center}
    \renewcommand{\arraystretch}{1.1} 
    \caption{Nominal parameter values}\label{tb:param}
        \begin{tabular}{ccccc}
	    	\cmidrule{1-2} \cmidrule{4-5}
            $ \ksp $ & $ 55 \,\mathrm{N/m} $ &&
            $ \kappa_5 $ & $1320\,\mathrm{m^{-1}}$\\
            $ \zsp $ & $ 0.015 \,\mathrm{m} $ &&
            $ \kappa_6 $ & $9.73\cdot10^{-3}\,\mathrm{m}$ \\
            $ m $ & $1.6 \cdot 10^{-3} \,\mathrm{kg} $ &&
            $ R $ & $ 50 \, \mathrm{\Omega}$\\
	    	$ \auxRelcz $ & $1.35 \,\mathrm{H^{-1}} $ &&
	    	$ z_0 $ & $10^{-3}\,\mathrm{m}$ \\
	    	$ \lambdasat $ & $0.0229 \,\mathrm{Wb} $ &&
	    	$ z_\mathrm{f} $ & $0$ \\
    		$ \auxRelgz  $ & $3.88 \,\mathrm{H^{-1}} $ &&
    		$ t_0 $ & $ 0 $ \\
            $ \auxdRelgz $ & $7.67  \cdot 10^{4} \,\mathrm{H^{-1}/m} $ &&
    		$ \tf $ & $ 3.5 \cdot 10^{-3}\,\mathrm{s}$ \\	
	    	\cmidrule{1-2} \cmidrule{4-5} 
        \end{tabular}
    \end{center}
    \vspace{-2mm}
\end{table}

\begin{table}[t]
    \begin{center}
    \renewcommand{\arraystretch}{1.1} 
    \caption{Initial hyperparameters of the control strategies}\label{tb:hyper}
        \begin{tabular}{ccccccc}
        Control    & $\s$ & $\s_\mathrm{min}$ & $\s_\mathrm{max}$ & $\alpha_\mathrm{con}$ & $\alpha_\mathrm{exp}$ & $\beta$ \\ \cmidrule{1-7}
        R2R-PS+ & $0.2$ & $2\cdot10^{-10}$ & $2$ & $0.5$ & $2$ & -- \vspace{1pt}\\
        R2R-AC & $0.2$ & $2\cdot10^{-10}$ & $2$ & $0.7$ & $1.1$ & $0.8$
        \end{tabular}
    \end{center}
    \vspace{-8mm}
\end{table}

In this section, to illustrate the benefits of the new algorithm, we analyze the improvements achieved over our previous work~\citep{edgar2024ECC}. To assess the improvements introduced by R2R-AC and the hybrid step size, the following control strategies are evaluated:  
\begin{itemize}
    \item \textbf{R2R-PS+}: the best solution in \citep{edgar2024ECC}; it uses for the iterative adapatation law a Pattern Search algorithm, with a single initial basis change \eqref{eq:theta_X} and four dimensions.
	\item \textbf{R2R-AC (DF)}: proposed R2R-AC in which $\deltaX \leftarrow \s$.
	\item \textbf{R2R-AC (GB)}: proposed R2R-AC in which $\deltaX \leftarrow \deltaX^\mathrm{GB}$.
	\item \textbf{R2R-AC (Hy)}: proposed R2R-AC in which $\deltaX$ is computed using the Algorithm{~\ref{alg:dx}}, the complete proposal.
\end{itemize}

These strategies have been tested through simulation on the problem presented in Section~\ref{sec:background}. Recall that the detailed methodology underlying the design of the feedforward controller, as well as the formulation of the desired trajectory, can be found in \mbox{\citet{edgar2024ECC}}. It is assumed that the model acting in the role of the real system and the feedforward controller are based on exactly the same equations. However, it is reasonable to expect discrepancies between the model parameters identification and the nominal (initial) feedforward controller parameters (see Table~\ref{tb:param}). To emulate these errors, two Monte Carlo analyses of 10\,000 trials, and 300 switching operations in each trial are performed for each run-to-run strategy. For each Monte Carlo analysis, a test set with a different \mbox{$\param$-vector}, \eqref{eq:param0}, for each trial is generated. The first test set is associated with a situation where the errors are small, and the second with a situation where the errors are larger. To ensure fair comparisons, these test sets are common for all algorithms. The initial hyperparameters required of each algorithm (see Table~{\ref{tb:hyper}}) have been adjusted to optimize the results of the test where the errors are small. The same hyperparameters have been applied to the other test.

\subsection{Small errors in the initial parameters}\label{sec:results_05}

\begin{figure}
    \subfloat[R2R-PS+. \label{fig:PS405}] {\includegraphics[trim={0 0 0 0.01mm},clip]{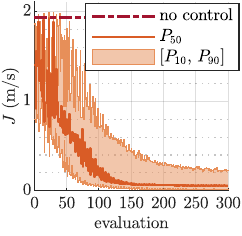}} \hspace{2mm}
	\subfloat[R2R-AC (DF). \label{fig:ACE-DF05}] {\includegraphics[trim={0 0 0 0.01mm},clip]{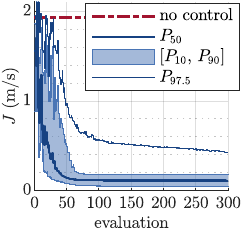}} \hspace{2mm}         \vspace{0mm} 
    \subfloat[R2R-AC (GB). \label{fig:ACE-GBb05}] {\includegraphics[trim={0 0 0 0.01mm},clip]{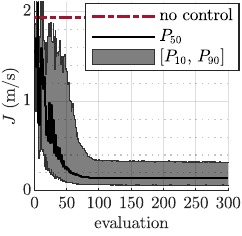}} \hspace{2mm}  
    \subfloat[R2R-AC (Hy). \label{fig:ACE-Hy05}] {\includegraphics[trim={0 0 0 0.01mm},clip]{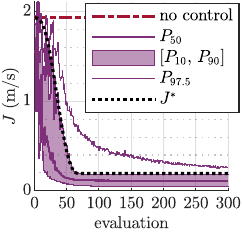}}%
     \vspace{0mm}
	\caption{Cost values with respect to the number of switching operations when parameter perturbations set to $5\,\%$. Each graph shows the median ($P_{50}$) and the 10th and 90th percentiles ($P_{10}$ and $P_{90}$, respectively) of the distribution of values obtained for the 10\,000 simulated experiments. The cost without control is also represented. The 97.5th percentil ($P_{97.5}$) is also represented in (b) and (d) to show the hybrid method improvement.}	\label{fig:results_05} \vspace{-0mm}
\end{figure}

\begin{figure}[t]   
	\subfloat[ \label{fig:Hy_slow}] {\includegraphics[trim={0 0 0 0.01mm},clip]{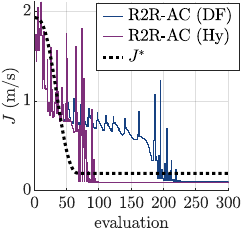}}\vspace{1mm}
	\subfloat[ \label{fig:Hy_conv}] {\includegraphics[trim={0 0 0 0.01mm},clip]{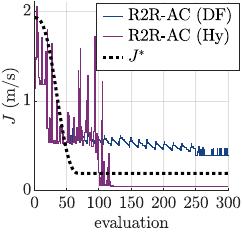}}%
  \vspace{-2mm}
	\caption{Effect of the Hybrid strategy versus Derivative-Free strategy. Evolution of $J$ in two specific processes selected as representative. (a) Process with slow convergence. (b) Process with convergence to an unacceptable cost}	 \vspace{-0mm} \label{fig:Hy_efect}
 \vspace{-5mm}
\end{figure}

For this situation, each component of each \mbox{$\param$-vector} of the real system model is randomly and independently perturbed up to $5\,\%$, i.e., the parameters of the real device under consideration vary with a uniform probability distribution between $95\,\%$ and $105\,\%$ of the values in Table~{\ref{tb:param}}.

Fig.~\ref{fig:results_05} shows the results of the first analysis. The graphs represent the evolution of the cost, $J$, with respect to each evaluation or switching operation. Due to the large number of simulations required to capture the variability of the parameters across devices, the results are presented by the median ($P_{50}$) and the 10th and 90th percentiles ($P_{10}$ and $P_{90}$, respectively) of the distribution of values obtained for the 10\,000 simulated experiments. For reference, the cost of a switching operation without control, namely with a 30~V constant activation, is also plotted. To demonstrate the improvement introduced by R2R-AC, Fig.~\ref{fig:PS405} (the best solution in~\citep{edgar2024ECC}) and~\ref{fig:ACE-DF05} must be compared, as they differ only in the search strategy; both methods compute the step size using the same derivative-free approach. As can be seen, while the results at the end are quite similar, the new \mbox{R2R-AC (DF)} strategy shows a notable improvement in the convergence speed. While our previous feedforward run-to-run controller requires 300 switching operations for 90\,\% of the trials to converge, \mbox{R2R-AC (DF)} requires only approximately 70. The same applies to 50\,\% and 10\,\% of the trials, for which the required number of switching operations is reduced by approximately half.

The other two strategies require the definition of $J^*$. For \mbox{R2R-AC (GB)}, since gradient-based methods are sensitive to excessively large step sizes, $J^*$ is variable and is calculated as the minimum evaluated cost, $J^\mathrm{min}$, multiplied by a positive constant $\gamma < 1$. With this strategy, the convergence speed also increases, but the final results are worse in this case. For \mbox{R2R-AC (Hy)}, the expression of $J^*$ (see Fig.~\ref{fig:ACE-Hy05}) is derived from a smoothed $P_{90}$ behavior of \mbox{R2R-AC (DF)} under the constraint that the initial value matches the cost obtained from an evaluation without control. In this way, processes below the 90th percentile should remain unchanged, as can be seen when Fig.~\ref{fig:ACE-DF05} and Fig.~\ref{fig:ACE-Hy05} are compared, while those above the 90th percentile should improve their performance when the hybrid method is applied to calculate the step size. For this reason, the 97.5th percentil ($P_{97.5}$) is also represented in these figures. Comparing this index, with \mbox{R2R-AC (Hy)} the target value is almost reached at the end of the trials, while with \mbox{R2R-AC (DF)} the convergence continues but progresses at a slow pace. Fig.~\ref{fig:Hy_slow} and Fig.~\ref{fig:Hy_conv} illustrate the effect of the hybrid strategy on the $J$ evolution of two individual processes, one with slow convergence and another that converges to an unacceptable cost, respectively.

\subsection{Larger errors in the initial parameters}\label{sec:results_25}
\begin{figure}[t]
    \subfloat[R2R-PS+. \label{fig:PS425}] {\includegraphics[trim={0 0 0 0.01mm},clip]{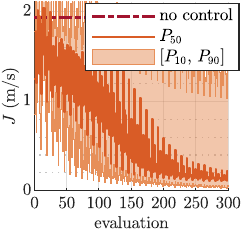}} \hspace{2mm}
	\subfloat[R2R-AC (DF). \label{fig:ACE-DF25}] {\includegraphics[trim={0 0 0 0.01mm},clip]{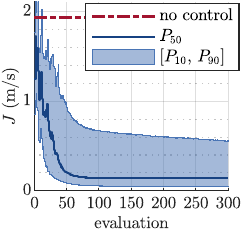}} \hspace{2mm}         \vspace{4mm} 
    \subfloat[R2R-AC (GB). \label{fig:ACE-GBb25}] {\includegraphics[trim={0 0 0 0.01mm},clip]{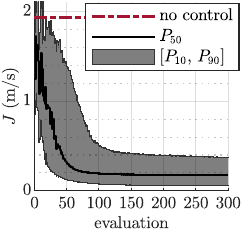}} \hspace{2mm}  
    \subfloat[R2R-AC (Hy). \label{fig:ACE-Hy25}] {\includegraphics[trim={0 0 0 0.01mm},clip]{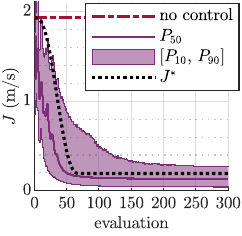}} \vspace{-2mm} %
	\caption{Cost values with respect to the number of switching operations when parameter perturbations set to $25\,\%$. Each graph shows the median ($P_{50}$) and the 10th and 90th percentiles ($P_{10}$ and $P_{90}$, respectively) of the distribution of values obtained for the 10\,000 simulated experiments. The cost without control is also represented.}	\label{fig:results_25} \vspace{-2mm}
\end{figure}

Analogous to the previous subsection, the $\param$ test set has been generated with perturbations up to $25\,\%$ instead of $5\,\%$.

Fig.~\ref{fig:results_25} shows the results of the second analysis. As in the previous case, $P_{10}$, $P_{50}$ and $P_{90}$ of the distribution of values obtained for the 10\,000 simulated experiments are shown. Each control strategy, i.e., \mbox{R2R-PS+}, \mbox{R2R-AC (DF)}, \mbox{R2R-AC (GB)}, and \mbox{R2R-AC (Hy)}, uses the same hyperparameters as those employed in the previous subsection. This includes the definition of $J^*$ across evaluations: for \mbox{R2R-AC (GB)}, \mbox{$J^* \leftarrow J^\mathrm{min} \cdot \gamma$}, and for \mbox{R2R-AC (Hy)}, $J^*$ is the smoothed behavior of $P_{90}$ with the set of nominal parameters perturbed up to 5\,\%.

As can be seen, the improvement of R2R-AC, compared to~\citep{edgar2024ECC} is considerable. In contrast to cases with initial low $\param$ error, \mbox{R2R-AC (GB)} is able to offer better performance than \mbox{R2R-AC (DF)}. However, our complete proposal, \mbox{R2R-AC (Hy)}, achieves the best results regardless of the error level. As a summary, Table~\ref{tb:sum_results} compares $P_{90}$ of $J$ of the 10\,000 simulated experiments at the 300th switching operation in both situations.

\begin{table}
    \begin{center}
    \setlength{\tabcolsep}{2pt} 
    \renewcommand{\arraystretch}{1.1} 
    \small
    \caption{Comparison of $P_{90}$ of $J~(\mathrm{m}/\mathbf{s})$ in the 300th switching operation}\label{tb:sum_results}
        \begin{tabular}{ccccc}
            Perturbation & R2R-PS+ & R2R-AC (DF) &  R2R-AC (GB) & \textbf{R2R-AC (Hy)} \\
	    	\cmidrule{1-5}
	    	5\,\%  & 0.2268 & 0.1777 & 0.3178 & \textbf{0.1772} \\
            \cmidrule{1-5}
            25\,\% & 1.2381 & 0.5541 & 0.3750 & \textbf{0.2645} \\
	    	\cmidrule{1-5} 
        \end{tabular}
    \end{center}
    \vspace{-7mm}
\end{table}

\section{Conclusions}\label{sec:conclusions}
In this work, we have presented \mbox{R2R-AC (Hy)}, a run-to-run control scheme with a new algorithm for iteratively adapting the parameters of a feedforward controller from indirect measurements. However, as outlined, the new algorithm could be used for other differentiably parametrized controllers. The improvement over the previous approach has been achieved both for small initial parameter errors and for larger errors, where the previous technique is not effective. The improvements have been obtained by integrating three concepts into the algorithm: a scheduled basis change based on the sensitivity of the feedforward law, a continuous process of exploration and exploitation, and a hybrid method that toggles between a derivative-free and a gradient-based strategy to calculate the step size. Likewise, this new algorithm automates two aspects of our previous work: the update of the coordinate system and the number of search dimensions to reduce, as the new algorithm selects the minimum number of dimensions to improve its feedforward controller behavior. 

As future work, we would like to address the possibility of an estimation technique for the objective function $J^*$ or to analyze and improve the gradient-based step-size calculation in other scenarios, such as stochastic processes or noisy measurements, by considering alternative approaches from the literature. In addition, we also intend to perform real laboratory tests on different systems to verify that the experimental results agree with those observed in simulation and the generality of the method.


\begin{thebibliography}{19}
\expandafter\ifx\csname natexlab\endcsname\relax\def\natexlab#1{#1}\fi
\providecommand{\url}[1]{\texttt{#1}}
\providecommand{\href}[2]{#2}
\providecommand{\path}[1]{#1}
\providecommand{\DOIprefix}{doi:}
\providecommand{\ArXivprefix}{arXiv:}
\providecommand{\URLprefix}{URL: }
\providecommand{\Pubmedprefix}{pmid:}
\providecommand{\doi}[1]{\href{http://dx.doi.org/#1}{\path{#1}}}
\providecommand{\Pubmed}[1]{\href{pmid:#1}{\path{#1}}}
\providecommand{\bibinfo}[2]{#2}
\ifx\xfnm\relax \def\xfnm[#1]{\unskip,\space#1}\fi
\bibitem[{Al~Saaideh et~al.(2022)Al~Saaideh, Boker and Al~Janaideh}]{mohammad2022output}
\bibinfo{author}{Al~Saaideh, M.}, \bibinfo{author}{Boker, A.M.}, \bibinfo{author}{Al~Janaideh, M.}, \bibinfo{year}{2022}.
\newblock \bibinfo{title}{Output-feedback control of electromagnetic actuated micropositioning system with uncertain nonlinearities and unknown gap variation}, in: \bibinfo{booktitle}{IEEE Conf. Decision and Control}, \bibinfo{organization}{IEEE}. pp. \bibinfo{pages}{2481--2486}.
\bibitem[{Angadi and Jackson(2022)}]{angadi2022critical}
\bibinfo{author}{Angadi, S.V.}, \bibinfo{author}{Jackson, R.L.}, \bibinfo{year}{2022}.
\newblock \bibinfo{title}{A critical review on the solenoid valve reliability, performance and remaining useful life including its industrial applications}.
\newblock \bibinfo{journal}{Eng. Fail. Anal.} \bibinfo{volume}{136}, \bibinfo{pages}{106231}.
\bibitem[{Benosman and At{\i}n{\c{c}}(2015)}]{Benosman2015}
\bibinfo{author}{Benosman, M.}, \bibinfo{author}{At{\i}n{\c{c}}, G.M.}, \bibinfo{year}{2015}.
\newblock \bibinfo{title}{{Extremum seeking-based adaptive control for electromagnetic actuators}}.
\newblock \bibinfo{journal}{Int. J. Control} \bibinfo{volume}{88}, \bibinfo{pages}{517--530}.
\bibitem[{Braun et~al.(2018)Braun, Reuter and Rudolph}]{braun2018observer}
\bibinfo{author}{Braun, T.}, \bibinfo{author}{Reuter, J.}, \bibinfo{author}{Rudolph, J.}, \bibinfo{year}{2018}.
\newblock \bibinfo{title}{Observer design for self-sensing of solenoid actuators with application to soft landing}.
\newblock \bibinfo{journal}{IEEE Trans. Control Syst. Technol.} \bibinfo{volume}{27}, \bibinfo{pages}{1720--1727}.
\bibitem[{Deschaux et~al.(2018)Deschaux, Gouaisbaut and Ariba}]{deschaux2018nonlinear}
\bibinfo{author}{Deschaux, F.}, \bibinfo{author}{Gouaisbaut, F.}, \bibinfo{author}{Ariba, Y.}, \bibinfo{year}{2018}.
\newblock \bibinfo{title}{Nonlinear control for an uncertain electromagnetic actuator}, in: \bibinfo{booktitle}{IEEE Conf. Decision and Control}, \bibinfo{organization}{IEEE}. pp. \bibinfo{pages}{2316--2321}.
\bibitem[{Gergič and Hercog(2019)}]{bojan2019design}
\bibinfo{author}{Gergič, B.}, \bibinfo{author}{Hercog, D.}, \bibinfo{year}{2019}.
\newblock \bibinfo{title}{Design and implementation of a measurement system for high-speed testing of electromechanical relays}.
\newblock \bibinfo{journal}{Measurement} \bibinfo{volume}{135}, \bibinfo{pages}{112--121}.
\bibitem[{Grotjahn and Heimann(2002)}]{grotjahn2002model}
\bibinfo{author}{Grotjahn, M.}, \bibinfo{author}{Heimann, B.}, \bibinfo{year}{2002}.
\newblock \bibinfo{title}{Model-based feedforward control in industrial robotics}.
\newblock \bibinfo{journal}{Int. J. Robot. Res.} \bibinfo{volume}{21}, \bibinfo{pages}{45--60}.
\bibitem[{Hansen and Ostermeier(2001)}]{hansen2001completely}
\bibinfo{author}{Hansen, N.}, \bibinfo{author}{Ostermeier, A.}, \bibinfo{year}{2001}.
\newblock \bibinfo{title}{Completely derandomized self-adaptation in evolution strategies}.
\newblock \bibinfo{journal}{Evol. Comput.} \bibinfo{volume}{9}, \bibinfo{pages}{159--195}.
\bibitem[{L{\'e}vine(2011)}]{levine2011}
\bibinfo{author}{L{\'e}vine, J.}, \bibinfo{year}{2011}.
\newblock \bibinfo{title}{On necessary and sufficient conditions for differential flatness}.
\newblock \bibinfo{journal}{Appl. Algebra Eng., Commun. Comput.} \bibinfo{volume}{22}, \bibinfo{pages}{47--90}.
\bibitem[{Lewis and Torczon(2000)}]{lewis2000pattern}
\bibinfo{author}{Lewis, R.M.}, \bibinfo{author}{Torczon, V.}, \bibinfo{year}{2000}.
\newblock \bibinfo{title}{Pattern search methods for linearly constrained minimization}.
\newblock \bibinfo{journal}{SIAM J. Optimization} \bibinfo{volume}{10}, \bibinfo{pages}{917--941}.
\bibitem[{Loshchilov et~al.(2011)Loshchilov, Schoenauer and Sebag}]{loshchilov2011adaptive}
\bibinfo{author}{Loshchilov, I.}, \bibinfo{author}{Schoenauer, M.}, \bibinfo{author}{Sebag, M.}, \bibinfo{year}{2011}.
\newblock \bibinfo{title}{Adaptive coordinate descent}, in: \bibinfo{booktitle}{Prod. 13th GECCO}, pp. \bibinfo{pages}{885--892}.
\bibitem[{Ma et~al.(2022)Ma, Wu and Weinan}]{ma2022qualitative}
\bibinfo{author}{Ma, C.}, \bibinfo{author}{Wu, L.}, \bibinfo{author}{Weinan, E.}, \bibinfo{year}{2022}.
\newblock \bibinfo{title}{A qualitative study of the dynamic behavior for adaptive gradient algorithms}, in: \bibinfo{booktitle}{Math. Sci. Mach. Learning}, \bibinfo{organization}{PMLR}. pp. \bibinfo{pages}{671--692}.
\bibitem[{Moulay et~al.(2019)Moulay, L{\'e}chapp{\'e} and Plestan}]{moulay2019properties}
\bibinfo{author}{Moulay, E.}, \bibinfo{author}{L{\'e}chapp{\'e}, V.}, \bibinfo{author}{Plestan, F.}, \bibinfo{year}{2019}.
\newblock \bibinfo{title}{Properties of the sign gradient descent algorithms}.
\newblock \bibinfo{journal}{Inf. Sci.} \bibinfo{volume}{492}, \bibinfo{pages}{29--39}.
\bibitem[{Moya-Lasheras et~al.(2023)Moya-Lasheras, Ramirez-Laboreo and Serrano-Seco}]{moya2023IFAC}
\bibinfo{author}{Moya-Lasheras, E.}, \bibinfo{author}{Ramirez-Laboreo, E.}, \bibinfo{author}{Serrano-Seco, E.}, \bibinfo{year}{2023}.
\newblock \bibinfo{title}{{Run-to-Run Adaptive Nonlinear Feedforward Control of Electromechanical Switching Devices}}.
\newblock \bibinfo{journal}{IFAC-PapersOnLine} \bibinfo{volume}{56}, \bibinfo{pages}{5358--5363}.
\newblock \bibinfo{note}{22nd IFAC World Congr.}
\bibitem[{Moya-Lasheras and Sagues(2024)}]{moya2024iterative}
\bibinfo{author}{Moya-Lasheras, E.}, \bibinfo{author}{Sagues, C.}, \bibinfo{year}{2024}.
\newblock \bibinfo{title}{Iterative-only learning control for soft landing of short-stroke reluctance actuators}.
\newblock \bibinfo{journal}{Control Eng. Pract.} \bibinfo{volume}{152}, \bibinfo{pages}{106067}.
\bibitem[{Ramirez-Laboreo et~al.(2024)Ramirez-Laboreo, Moya-Lasheras and Serrano-Seco}]{edgar2024ECC}
\bibinfo{author}{Ramirez-Laboreo, E.}, \bibinfo{author}{Moya-Lasheras, E.}, \bibinfo{author}{Serrano-Seco, E.}, \bibinfo{year}{2024}.
\newblock \bibinfo{title}{Faster run-to-run feedforward control of electromechanical switching devices: A sensitivity-based approach}, in: \bibinfo{booktitle}{2024 European Control Conf.}, pp. \bibinfo{pages}{1321--1326}.
\bibitem[{Schroedter et~al.(2018)Schroedter, Roth, Janschek and Sandner}]{schroedter2018flatness}
\bibinfo{author}{Schroedter, R.}, \bibinfo{author}{Roth, M.}, \bibinfo{author}{Janschek, K.}, \bibinfo{author}{Sandner, T.}, \bibinfo{year}{2018}.
\newblock \bibinfo{title}{Flatness-based open-loop and closed-loop control for electrostatic quasi-static microscanners using jerk-limited trajectory design}.
\newblock \bibinfo{journal}{Mechatronics} \bibinfo{volume}{56}, \bibinfo{pages}{318--331}.
\bibitem[{Winkel et~al.(2023)Winkel, Scholz, Wallscheid and B{\"o}cker}]{winkel2023reducing}
\bibinfo{author}{Winkel, F.}, \bibinfo{author}{Scholz, P.}, \bibinfo{author}{Wallscheid, O.}, \bibinfo{author}{B{\"o}cker, J.}, \bibinfo{year}{2023}.
\newblock \bibinfo{title}{Reducing contact bouncing of a relay by optimizing the switch signal during run-time}.
\newblock \bibinfo{journal}{IEEE Trans. Autom. Sci. Eng.} .
\bibitem[{Yeh and Hsu(1999)}]{yeh1999optimal}
\bibinfo{author}{Yeh, S.S.}, \bibinfo{author}{Hsu, P.L.}, \bibinfo{year}{1999}.
\newblock \bibinfo{title}{An optimal and adaptive design of the feedforward motion controller}.
\newblock \bibinfo{journal}{IEEE/ASME Trans. Mechatronics} \bibinfo{volume}{4}, \bibinfo{pages}{428--439}.

\end{thebibliography}
\end{document}